\documentstyle[11pt,psfig,subfigure]{article}

\begin{document}

\hfill    SU-4240-597

\hfill November 1994

\begin{center}

\vspace{24pt}
{\Large \bf The forgotten Potts models}
\vspace{24pt}
   
{\large \sl Gudmar Thorleifsson}

\vspace{6pt}

Department of Physics \\ 
Syracuse University \\
Syracuse, NY 12344-1130, USA 
        
\vspace{12pt}
\begin{abstract}
The $q=10$ and $q=200$ state Potts models coupled to
$2d$ gravity are investigated numerically and 
shown to have continuous phase transitions, contrary to
their behavior on a regular lattice.
Critical exponents are extracted and possible critical
behavior for general $q$-state Potts models coupled
to gravity is discussed. 

\end{abstract}
\end{center}

\section{Introduction}

In the continuum formulation of $2d$ quantum gravity one is
interested in studying the coupling of conformally
invariant matter to gravity.  The conformal
matter is characterized by its central charge $c$ but
yet it has only been possible to solve theories with
$c \leq 1$ analytically. 

This has triggered a lot of numerical work on 
discretized models of $2d$ gravity with $c>1$ \cite{r1}. 
One such
discretization method is dynamical triangulations
where the integration over space-time manifolds is 
replaced by a sum over piecewise linear surfaces constructed
by gluing together equilateral triangles.  The matter
is simulated by spins, living either on triangles or
vertices.  The spin models usually used 
are the $q=2$, 3 and 4-state Potts models and the
Gaussian model.  All these models have continuous phase
transitions on regular lattices, with divergent
correlation
length, and hence can be associated  
with some conformal field theories in the continuum limit.
The above mentioned models all correspond to conformal
theories with $c\leq 1$ but taking multiple copies of them
allows one to study numerically theories with $c>1$ coupled
to gravity. For unlike in the continuum formulation the
discretized models are well defined for $c>1$.

For $q>4$ the Potts models have a discontinuous
or 1$st$ order phase transition on regular lattices
and hence no corresponding conformal field theories.
So until now coupling them to gravity has been 
discarded as uninteresting.  But a recent mean field
solution of the $q=\infty$ state Potts model coupled
to gravity showed the model to have a 3$rd$ order 
magnetization transition from a high-temperature
branched polymer phase to a low temperature
pure gravity phase ($c=0$) \cite{r2}. Moreover the model was shown 
to be equivalent to infinite copies of Ising models coupled
to gravity (a $c=\infty$ model).   

By expanding around the $q=\infty$ solution this
analysis was extended down to finite values of $q$.
There {\it two} phase transitions were found with 
high- and low-temperature pure gravity phases
separated by  
a branched polymer phase. The critical behavior of those
transitions was identical, both had the specific heat
exponent  
$\alpha = -1$. As the value of
$q$ is decreased the two phase transitions merge 
at some critical value $q_c$ 
and the branched polymer phase disappears.  At that value
of $q$ (estimated to be $q_c \approx 120$ in \cite{r2})
the perturbation expansion breaks down. 

It is worth comparing this with the situation for multiple
$q\leq 4$ state Potts models coupled to gravity. 
For $c\leq 1$ the interaction
between gravity and matter is weak, for all values of the
coupling {\it except} $\beta_c$ we have a pure gravity
phase. But for larger values of $c$ 
numerical simulations
indicate a high- and low-temperature pure gravity phase
with a branched polymer phase in between, i.e. the same
scenario as for high-$q$.

So the similarity between these two cases might indicate
that the study of Potts models for large value
of $q$ might give us informations about models with
large central charge coupled to gravity.
It might be possible that after coupling to
gravity these model would describe some conformal
matter coupled to gravity, the only problem being that
we have not yet identified these conformal theories.

This has prompted us to look at the $q=10$ and $q=200$ state
Potts models coupled to gravity applying 
Monte Carlo simulations     
\cite{r3}.  
Using micro-canonical simulations (fixed
area) the models are defined by 
\begin{equation}
Z(\beta,N) = \sum_{T\in {\cal T}(N)} \sum_{\{ \sigma_i \}}
{\rm exp} \left ( \beta \sum_{(i,j)} \delta_{\sigma_i,\sigma_j}
\right )
\end{equation}
where $\sigma_i \in \{1,...,q\}$ are the Potts spins, $i$
is a lattice triangle, $\{ \sigma_i \}$ a spin configuration
on the triangulation $T$ and ${\cal T}(N)$ an appropriate
class of triangulations (of size $N$). In these simulations
we include 
triangulations allowing  
sites joined by more than
one link and sites connected to them selves as 
these can be shown to reduce finite size effects \cite{r3}. 
The triangulations were updated using the link-flip
algorithm and the spin models with the
Swendsen-Wang cluster algorithm.
The lattice
sizes used were between 250 and 8000 triangles and, as the
auto-correlations increase linearly with $q$, 
we used $5 \times 10^6$
and $2 \times 10^7$ sweeps for each $\beta$ value for the 
$q=10$ and $q=200$ state Potts model respectively. 

\begin{figure}[t]
\begin{center}
\subfigure{
\psfig{figure=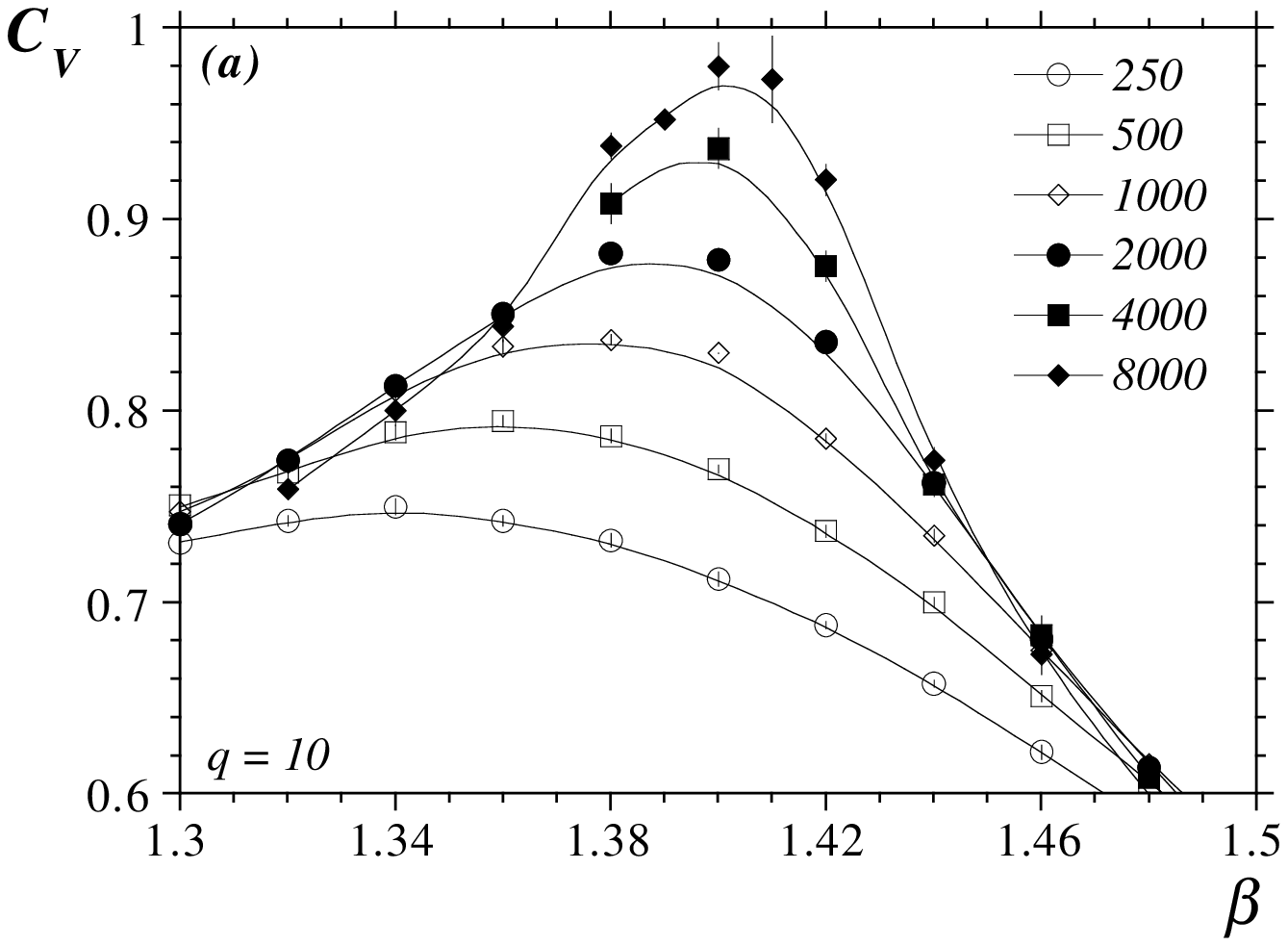,width=2.35in} }
\subfigure{
\psfig{figure=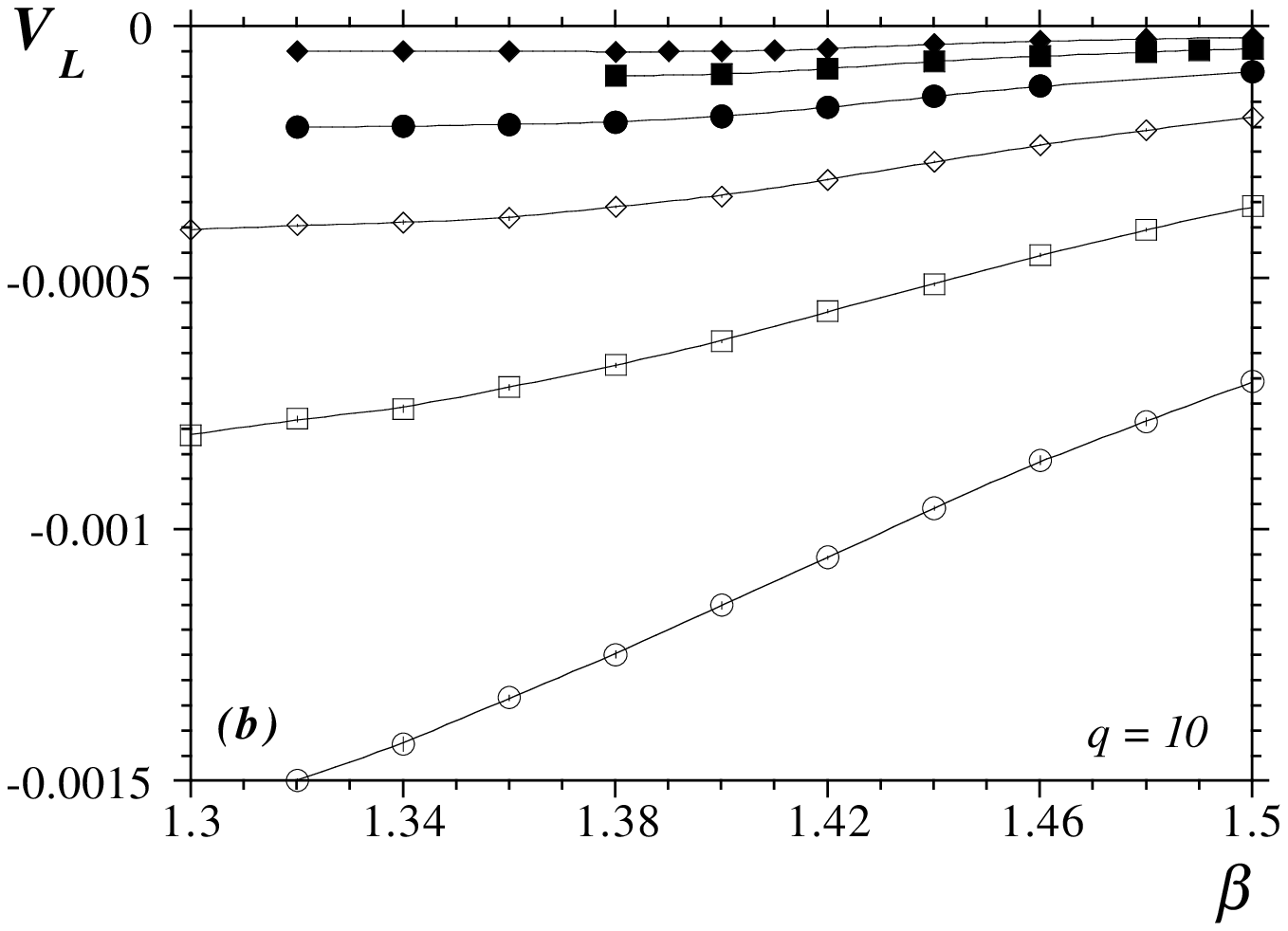,width=2.35in} }
\end{center}
\caption{(a) The specific heat $C_V$ versus $\beta$,
for different lattice sizes, for the
$q=10$ state Potts model coupled to gravity.(b)
The energy cumulant $V_L$ versus $\beta$ for the
same model}
\label{fig:largenenough}
\end{figure}

\section{The order of the phase transitions}

The first interesting question is the order of the phase
transitions, are they continuous or discontinuous after coupling
to gravity. 
This we determine by looking at the scaling 
of the maximum of the
specific heat with lattice size. If a phase
transition is 1{\it st} order $C_V(max)$ should scale
linearly with the volume.
From fig. 1{\it a}, where we plot the specific heat versus 
$\beta$ for the $q=10$ state Potts model, we
see that the maximum increases slowly with lattice size.
Looking at the quantity $C_V(max))/N$
it approaches zero in the infinity volume
limit, which is a strong indication for a continuous
phase transition.

Another quantity sensitive to the order of the
phase transition is the energy cumulant 
\begin{equation}
V_L = 1 - \frac{ <E^4>}{<E^2>^2 }.
\end{equation}
If the transition is 1{\it st} order $V_L$ should reach
a non-zero minimum in the critical point, 
otherwise it is zero for all values of $\beta$.
In fig. 1{\it b} we show that for the $q=10$ state Potts
model no indication of a non-zero minimum
can be seen.

Exactly the same is seen for the $q=200$ state
Potts model, and
from these measurements we conclude that 
both models have
continuous magnetization transitions after coupling
to gravity.
It it might at first seem strange that the order
of the phase transition can change that drastically
with coupling to
gravity, but we should remember that 
even for one Ising model (which can
be solved explicitly) there is a change from a 2{\it nd}
to a 3{\it rd} order transition with coupling to gravity.  

\section{Critical exponents}

\begin{table*}[t]
\begin{center}
\begin{tabular}{ccc} 
$q$ & $\beta_c(C_V)$ & $\beta_c(BC)$ \\
\hline
10 & 1.412(1)  & 1.409(2)  \\
200 & 2.633(1)  & 2.629(3) \\
\hline
\end{tabular}

\vspace{12pt}
\begin{tabular}{ccccccc}
$q$ & $\nu d^{(a)}$ & $\nu d^{(b)}$ & 
$\beta$ & $\gamma$ & $\alpha^{(a)}$ &
$\alpha^{(b)}$ \\
\hline
10  & 2.01(5)  & 2.03(4)  & 0.53(1)
& 1.11(2)  & 0.02(4)  & 0.02(3)  \\
200 & 2.49(15) & 2.52(12) & 1.18(4)
& 0.31(5)  & -0.83(6) & -0.80(5)  \\
\hline
\end{tabular}
\end{center}
\caption{Measured critical exponents for the $q=10$ and
$q=200$ state Potts models coupled to gravity.
For $\nu d_H$ we have values from {\it (a)} the scaling of
$\max \{\partial BC/\partial \beta\}$
and {\it (b)} the scaling of $M$ and
$\partial M/ \partial \beta$. In a same way for $\alpha$
{\it (a)} comes from $C_V(max)$ and
{\it (b)} from $C_V(\beta_c)$.}
\label{tab:tab1}
\end{table*}

To locate the critical temperatures for the
models we used the finite size scaling behavior
of two quantities: {\it (a)}
location of the peak in the specific heat
and {\it (b)} the intersection of Binder's cumulant
for different lattice sizes.  Both quantities are 
expected to approach $\beta_c$ as $N^{-1/\nu d_H}$.
The fits to the scaling behavior are made easier
as $\nu d_H$ can be determined directly from 
Binder's cumulant, as the maximum of its slope
scales as $N^{\nu d_H}$.
The results are shown in
table 1. 
  
We then applied standard
finite size scaling in $\beta_c$ to extract the
the critical exponents $\beta$ (from the
magnetization $M$), $\nu d_H$ (from $\partial M/ \partial
\beta$), $\gamma$ (from the magnetic susceptibility $\chi$)
and $\alpha$ (both from the scaling of $C_V(max)$ and
$C_V(\beta_c)$).  The results are shown in table 1.
For $q=10$ we tested that the scaling of the specific heat
fits equally well to logarithmic divergence
($C_V \sim \log(N)$).

\section{Discussion}

It is an interesting observation that the critical
exponents for the $q=10$ state Potts model 
are very close to that of
the $q=4$ state Potts model coupled to gravity
(which has $\beta = 1/2$,
$\gamma = 1$, $\alpha = 0$ and $\nu d_H = 2$).
As the value $q=10$ is chosen arbitrary this might
indicate that there exists an whole range of $q$ values,
starting at $q=4$ and ending at some $q_c$, where we
have the same critical behavior after coupling to
gravity.

On the other hand the $q=200$ state Potts model 
has $\alpha$ close to $-1$, so we might be
in the high-$q$ region were the calculation 
in \cite{r2} is valid. 
Then we would expect a branched polymer phase and two phase 
transitions.
Unfortunately we have only been able to locate one,
but as the other one is not a magnetization
transition it is not clear how to identify it.
But for an interval of $\beta$, before $\beta_c$, we
see some evidence of a branched polymer phase, supporting
the picture in \cite{r2}.

From this the following
critical behavior of the $q$-state Potts models
coupled to gravity sounds reasonable:  
For $q=4$ up to some $q_{c_1}$
we have the same critical behavior, i.e. a 2{\it nd}
order phase transition ($\alpha=0$) where the details
of the Potts models are less important than the coupling
to gravity.  And for 
$q > q_{c_2}$ we also have the same
critical behavior, this time with (presumable two)
3{\it rd} order phase transitions ($\alpha = -1$)
and a branched polymer phase.  
Whether $q_{c_1} = q_{c_2}$ or if we have some
different behavior at intermediary values of $q$
is not determined in these simulations.
But it is clear that we need some theoretical understanding
of what kind of conformal matter (if any) 
the $q > 4$ state Potts model coupled to gravity describes.

\end{document}